Supporting Information

# Physical properties of $ReB_2$ under high pressure and effect of metallic bonding on its hardness

MING-MIN ZHONG[1], XIAO-YU KUANG[1(a)], ZhEN-HUA WANG[1], PENG SHAO[1], LI-PING DING[1] and XIAO-FEN HUANG[2]

[1] *Institute of Atomic and Molecular Physics, Sichuan University, Chengdu 610065, China*

[2]*Physics Department, Sichuan Normal University, Chengdu 610068, China*

In this paper, we perform first-principles calculations to investigate the lattice parameters, total energies and mechanical stability of the $ReB_2$ in eight possible crystal structures. These eight potential $ReB_2$ structures based on the known transition metal and light element compounds are determined: cubic pyrite structure (*Pa-3*, p-$ReB_2$), simple tetragonal structure (*P4/mbm*, ST-$ReB_2$), simple hexagonal structure (*P6/mmm*, SH-$ReB_2$), hexagonal rhenium diboride structure (*P63/mmc* $ReB_2$-$ReB_2$), hexagonal aluminum diboride structure (*P6/mmm*, $AlB_2$-$ReB_2$), orthorhombic osmium diboride structure (*Pmmnz*, $OsB_2$-$ReB_2$), orthorhombic marcasite structure (*Pnnm*, m-$ReB_2$) and trigonal molybdenum diboride structure (*R-3m*, $MoB_2$-$ReB_2$).

After full geometry optimizations, all $ReB_2$ structures keep the same symmetry as the initial polymorphs. To further verify the mechanical stability of the eight polymorphs, their elastic constants are calculated. The elastic constants of a given crystal should satisfy the generalized elastic stability criteria [1]. For cubic crystal, $C_{11} > 0$, $C_{44} > 0$, $C_{11} - C_{12} > 0$, $C_{11} + 2C_{12} > 0$; for trigonal crystal, $C_{11} - C_{12} > 0$, $[C_{44}(C_{11} - C_{12}) - 2C_{14}^2] > 0$, $[C_{33}(C_{11} + C_{12}) - 2C_{13}^2] > 0$; for tetragonal crystal, $C_{11} > 0$, $C_{33} > 0$, $C_{44} > 0$, $C_{66} > 0$, $C_{11} - C_{12} > 0$, $C_{11} + C_{33} - 2C_{13} > 0$, $[2(C_{11} + C_{12}) + C_{33} + 4C_{13}] > 0$; for hexagonal crystal, $C_{44} > 0$, $C_{11} - C_{12} > 0$, $[C_{33}(C_{11} + C_{12}) - 2C_{13}^2] > 0$; and for orthorhombic crystal $C_{ii} > 0$, (i = 1, 2…6), $[C_{11} + C_{22} + C_{33} + 2(C_{12} + C_{13} + C_{23})] > 0$, $C_{11} + C_{22} - 2C_{12} > 0$, $C_{11} + C_{33} - 2C_{13} > 0$, $C_{22} + C_{33} - 2C_{23} > 0$. According to the mechanical stability criteria, the p-$ReB_2$, ST-$ReB_2$, SH-$ReB_2$, $AlB_2$-$ReB_2$, and m-$ReB_2$ are unstable, and their equilibrium lattice parameters and elastic constants are listed in Table S1 (in Supporting Information)

Table S1: Calculated lattice parameters (Å), elastic constants (GPa), cell volume per formula unit $V_0$ (Å$^3$), elastic constant (GPa).

| | p-$ReB_2$ | ST-$ReB_2$ | SH-$ReB_2$ | $AlB_2$-$ReB_2$ | m-$ReB_2$ |
|---|---|---|---|---|---|
| $a$ | 5.080 | 4.331 | 2.609 | 2.911 | 4.807 |
| $b$ | | | | | 4.149 |
| $c$ | | 2.969 | 5.957 | 3.453 | 3.029 |
| $V_0$ | 32.78 | 27.85 | 35.12 | 25.35 | 30.20 |
| $C_{11}$ | -345 | 238 | 197 | 678 | 478 |
| $C_{22}$ | | | | | 551 |
| $C_{33}$ | | 477 | 781 | 603 | 320 |
| $C_{44}$ | 164 | 197 | -70 | -18 | -131 |
| $C_{55}$ | | | | | 97 |
| $C_{66}$ | | 51 | -64 | 232 | -646 |
| $C_{12}$ | 494 | 303 | 324 | 214 | 127 |
| $C_{13}$ | | 172 | 25 | 173 | 254 |
| $C_{23}$ | | | | | 158 |

Table S2: Calculated bond parameters and Vickers hardness of transition metal monocarbides and mononitrides.

| Crystal | $v_b$ (Å$^3$) | $P$ | $f_m$ ($10^{-3}$) | $H_{v.\ \mathrm{calc.}}$ | $H_{v.\ \mathrm{expt.}}$ [a] |
|---|---|---|---|---|---|
| CrN | 2.741 | 0.335 | 6.997 | 11.3 | 11.0 |
| TaN | 3.539 | 0.360 | 1.315 | 27.7 | 22.0 |
| VC | 2.965 | 0.380 | 3.350 | 28.2 | 29.0 |
| HfC | 4.293 | 0.408 | 1.089 | 23.3 | 25.5 |
| TaC | 3.613 | 0.410 | 1.639 | 29.2 | 29.0 |
| WC | 3.434 | 0.400 | 2.434 | 27.5 | 30.0 |
| TiC | 3.353 | 0.333 | 1.113 | 28.7 | 29.0 |
| ZrC | 4.297 | 0.350 | 1.162 | 19.8 | 25.8 |
| TiN | 3.167 | 0.278 | 2.297 | 22.4 | 23.0 |
| ZrN | 3.994 | 0.270 | 1.596 | 16.4 | 15.0 |
| HfN | 4.131 | 0.325 | 1.567 | 18.7 | 17.0 |
| VN | 2.944 | 0.262 | 4.843 | 14.1 | 15.0 |
| NbN | 3.567 | 0.247 | 2.503 | 15.8 | 14.0 |

[a]Reference [2].

Table S3: Calculated bond parameters and Vickers hardness of crystals with a zinc blende structure.

| Crystal | $v_b$ (Å$^3$) | $P$ | $H_{v.\ \mathrm{calc.}}$ | $H_{v.\ \mathrm{expt.}}$[a] |
|---|---|---|---|---|
| Diamond | 2.755 | 0.75 | 96.8 | 96 |
| Si | 9.743 | 0.74 | 11.6 | 12 |
| Ge | 10.679 | 0.54 | 7.3 | 8.8 |
| Sn | 16.293 | 0.65 | 4.3 | 4.5 |
| SiC | 4.961 | 0.69 | 33.4 | 31 |
| BN | 2.888 | 0.65 | 77.6 | 63 |
| BP | 5.591 | 0.75 | 29.8 | 33 |
| BAs | 6.429 | 0.73 | 22.9 | 19 |
| AlP | 9.934 | 0.63 | 9.6 | 9.4 |
| AlAs | 11.090 | 0.61 | 7.7 | 5 |
| AlSb | 13.989 | 0.65 | 5.6 | 4 |
| GaP | 10.198 | 0.62 | 9.0 | 9.5 |
| GaSb | 13.995 | 0.47 | 4.0 | 4.5 |
| InP | 13.035 | 0.56 | 5.4 | 5.4 |
| InAs | 14.374 | 0.51 | 4.2 | 3.8 |
| InSb | 17.194 | 0.56 | 3.4 | 2.2 |

[a]Reference [2].